\documentclass[aps,superscriptaddress,eqsecnum,nofootinbib,showpacs,preprintnumbers,twocolumn]{revtex4}
\pdfoutput=1
\usepackage{xcolor}
\usepackage{longtable}
\usepackage{bm}
\usepackage{relsize}
\usepackage{amsfonts}
\usepackage{amsmath}
\usepackage{amssymb,epsf}
\usepackage{latexsym}
\usepackage{graphicx,epsfig}
\usepackage{amssymb}
\usepackage{float}
\usepackage{subfigure}
\usepackage{epstopdf}
\usepackage[colorlinks=true,citecolor=blue,linkcolor=blue,urlcolor=black]{hyperref}
\usepackage{dcolumn}
\usepackage{psfrag}
\usepackage{wrapfig}
\usepackage{makeidx}
\usepackage{epsf}
\usepackage{color}
\usepackage{multirow}
\usepackage{mathtools}

\begin{document}

\title{Gravitational wave propagation in generalized hybrid metric-Palatini gravity}

\author{Cláudio Gomes}
\email{claudio.gomes@fc.up.pt}
\affiliation{Centro de Física das Universidades do Minho e do Porto, Faculdade de Ciências da Universidade do Porto, Rua do Campo Alegre s/n, 4169-007 Porto, Portugal}
\affiliation{Universidade dos Açores, Instituto de Investigação em Ciências do Mar - OKEANOS, Campus da Horta, Rua Professor Doutor Frederico Machado 4, 9900-140 Horta, Portugal}

\author{João Luís Rosa}
\email{joaoluis92@gmail.com}
\affiliation{Departamento de F\'isica Te\'orica, Universidad Complutense de Madrid, E-28040 Madrid, Spain}
\affiliation{Institute of Physics, University of Tartu, W. Ostwaldi 1, 50411 Tartu, Estonia}

\author{Miguel A. S. Pinto}
\email{mapinto@fc.ul.pt}
\affiliation{Instituto de Astrofísica e Ciências do Espaço, Faculdade de Ciências da Universidade de Lisboa, Edifício C8, Campo Grande, P-1749-016 Lisbon, Portugal}
\affiliation{Departamento de F\'{i}sica, Faculdade de Ci\^{e}ncias da Universidade de Lisboa, Edifício C8, Campo Grande, P-1749-016 Lisbon, Portugal}

\date{\today}

\begin{abstract} 
In this work we analyze the propagation properties of gravitational waves in the generalized hybrid metric-Palatini gravity theory. We introduce the scalar-tensor representation of the theory to make explicit the scalar degrees of freedom of the theory and obtain their equations of motion in a form decoupled from the metric tensor. Then, we introduce linear perturbations for the metric tensor and for the two scalar fields and obtain the propagation equations for these three quantities. We analyzed the theory both at non-linear and at linear level through the Newman-Penrose formalism so to find the polarization states. We show that the tensor modes propagate at the speed of light and feature the usual $+$-- and $\times$--polarization modes also present in General Relativity (GR), plus two additional polarization modes: a longitudinal mode and a breathing mode, described by the same additional degree of freedom. On the other hand, the theory features two additional scalar modes not present in GR. These modes are massive and, thus, propagate with a speed smaller than the speed of light in general. The masses of the scalar modes depend solely on the interaction potential between the two scalar fields in the theory, which suggests that one can always fine-tune the potential to make the scalar modes massless and reduce their propagation speed to the speed of light. Given the possibility of fine-tuning the theory to match the observational predictions of GR and in the absence of any measured deviations, these features potentially render the generalized hybrid metric-Palatini theory unfalsifiable in the context of gravitational wave propagation. 
\end{abstract}

\pacs{04.50.Kd,04.20.Cv,}

\maketitle

\section{Introduction}\label{sec:intro}

The current cosmological measurements indicate that the universe is undergoing a phase of accelerated expansion \cite{SupernovaCosmologyProject:1998vns,SupernovaSearchTeam:1998fmf,Planck:2018vyg}. This behavior can be explained in the context of General Relativity (GR) as the effect of a relativistic perfect fluid with a negative pressure component, commonly denoted dark energy \cite{Copeland:2006wr,Li:2011sd,Peebles:2002gy}. However, an alternative explanation for this cosmological behavior can be attributed to modifications of the gravitational theory itself, known as modified theories of gravity \cite{Bamba:2012cp,Clifton:2011jh,Capozziello:2011et,Nojiri:2017ncd}. All types of modified theories of gravity seem to have their own advantages and disadvantages. One of the most popular families of models is the $f\left(R\right)$ gravity \cite{Sotiriou:2008rp,DeFelice:2010aj}, which successfully explains the accelerated cosmological expansion without invoking dark energy sources, as well as the dynamics of gravitating systems in the presence of dark matter \cite{Boehmer:2007kx}. However, the theory not only fails to reproduce the observed weak-field solar-system dynamics \cite{Khoury:2003aq,Khoury:2003rn,Dolgov:2003px,Sotiriou:2006sf}, but also raises other issues in a cosmological context \cite{Amendola:2006kh,Amendola:2006we,Tsujikawa:2007tg,delaCruz-Dombriz:2008ium}. A Palatini formulation of the $f\left(R\right)$ gravity was developed in an attempt to overcome these limitations \cite{Olmo:2011uz}, but unfortunately it still presents limitations in the context of compact stars \cite{Kainulainen:2006wz} and cosmological perturbations \cite{Koivisto:2005yc,Koivisto:2006ie}.

A combined version of both the metric and the Palatini formulations of $f\left(R\right)$ gravity, known as hybrid metric-Palatini gravity \cite{Capozziello:2015lza}, successfully addresses the problem of unifying the late-time cosmic acceleration phase with the weak-field solar system dynamics \cite{Harko:2011nh,Capozziello:2013uya}. Later on, the theory was proven useful to overcome a wide variety of limitations of $f\left(R\right)$ gravity in the fields of astrophysics and cosmology \cite{Capozziello:2012ny,Capozziello:2012qt,Koivisto:2013kwa,Capozziello:2012hr,Capozziello:2013yha,Carloni:2015bua,Edery:2019txq,Harko:2020ibn,Danila:2016lqx,Danila:2018xya,Bronnikov:2019ugl,Leanizbarrutia:2017xyd,Bekov:2020dww,Chen:2020evr,Dyadina:2023exu}. A generalization of the theory to allow for an arbitrary dependence of the action in the Ricci scalars of both the metric and the Palatini formalism was also developed \cite{Tamanini:2013ltp} (which is known as generalized hybrid metric-Palatini gravity, or GHMPG), leading to an even broader range of applications in these fields \cite{Rosa:2018jwp,Rosa:2017jld,Rosa:2020uoi,Rosa:2019ejh,Borowiec:2020lfx,Rosa:2021yym,Rosa:2020uli,Bombacigno:2019did,Rosa:2021mln,Sa:2020qfd,Rosa:2021lhc,Rosa:2021ish,daSilva:2021dsq,Bronnikov:2021tie,Golsanamlou:2023wiz,Rosa:2024pzo,Bombacigno:2024lud}. In this work, we aim to extend this literature to the field of gravitational waves (GWs).

The observation of GWs marked the beginning of a new era in fundamental physics research \cite{LIGOScientific:2018mvr,KAGRA:2023pio,Barack:2018yly,Bailes:2021}, providing remarkable insights into the nature of compact objects and binary systems \cite{Cardoso:2017cqb,Berti:2019wnn,LIGOScientific:2018jsj}, tests of gravity in the strong-field regime \cite{LIGOScientific:2016lio,Yunes:2016jcc,Berti:2018cxi,Berti:2018vdi}, and even cosmological measurements \cite{LIGOScientific:2017adf,DES:2019ccw}. Naturally, the propagation of gravitational waves was also analyzed in the context of modified theories of gravity \cite{Bernard:2022noq,Dalang:2020eaj,Capozziello:2006ra,Katsuragawa:2019uto,Gong:2017bru,Abedi:2017jqx,Capozziello:2008,Lopez:2025gfu,Casado-Turrion:2024esi}.
Interestingly however, only a few works have addressed this topic in hybrid metric-Palatini gravity \cite{Dyadina:2022elo,Dyadina:2024jjo,Kausar:2018ipo}. The results presented in these publications are shallow: the polarization modes are obtained from a purely perturbative approach, and none of the works take into consideration the generalized version of the theory. This work aims to suppress these gaps by extending the analysis to the generalized version of the theory and including an analysis of the polarization modes based on the Newman-Penrose (NP) formalism \cite{NP}. 

The NP formalism is based on a tetrad basis, and it is useful to analyze black hole and gravitational wave solutions. For metric theories of gravity that allow for plane null gravitational wave solutions, one can resort to the Eardley-Lee-Lightman (ELL) formalism which simplifies the former and allows for the classification of the spacetime solutions \cite{ELL1,ELL2}, based on which one can analyze all six polarization states a metric theory can exhibit. In fact, this formalism has been applied to modified gravity theories \cite{Alves:2009,Gomes:2018}. However, when a propagating mode is massive, the ELL formalism cannot be applied directly, and one has to invert its definitions to correctly find the polarization states \cite{Polarisations}. A notable exception arises if the mass of the polarization mode is small, for which one obtains nearly null plane gravitational waves and the ELL formalism is argued to be approximately applicable \cite{Capozziello:2020}. Moreover, there is a distinction between polarization modes that affect a ring of test particles when a gravitational wave passes through and the independent radiative degrees of freedom a theory may exhibit; hence, for instance, $f(R)$ theories possess three radiative degrees of freedom which can excite up to four polarization states \cite{Alves:2024}.

This manuscript is organized as follows. In Sec. \ref{sec:theory}, we introduce the GHMPG theory in both the geometrical and scalar-tensor representations and obtain their respective decoupled equations of motion; in Sec. \ref{sec:linear} we introduce linear perturbations of the metric and scalar fields to obtain the propagation equations and analyze the polarization modes of plane gravitational waves; in Sec. \ref{sec:newman} we introduce the Newman-Penrose formalism to determine the polarization states and degrees of freedom of gravitational waves when massive modes are present; in Sec. \ref{sec:implications} we analyze the results obtained from both methods above and their implications for the theory from an observational point of view; and in Sec. \ref{sec:concl} we discuss our results and trace our conclusions.

\section{Theory and equations}\label{sec:theory}

\subsection{Generalized hybrid metric-Palatini gravity}

The action $S$ that describes the GHMPG theory is given by
\begin{equation}\label{eq:gaction}
S=\frac{1}{2\kappa^2}\int_\Omega\sqrt{-g}f\left(R,\mathcal R\right)d^4x+S_m,
\end{equation}
where $\kappa^2\equiv 8\pi G/c^4$, with $G$ is the gravitational constant and $c$ is the speed of light, both set to $G=c=1$ through the introduction of a geometrized unit system such that $\kappa^2=8\pi$, $\Omega$ is the spacetime manifold, $g$ is the determinant of the spacetime metric $g_{\mu\nu}$ written in the system fo coordinates $x^\mu$, $f\left(R,\mathcal R\right)$ is a well-behaved function of the following two geometrical scalars, $R\equiv g^{\mu\nu} R_{\mu\nu}$, which is the Ricci scalar of the metric $g_{\mu\nu}$ where $R_{\mu\nu}$ is the Ricci tensor constructed in terms of the Levi-Civita connection $\Gamma^\alpha_{\mu\nu}$, and $\mathcal R\equiv g^{\mu\nu}\mathcal R_{\mu\nu}$ which is the Palatini Ricci scalar where $\mathcal R_{\mu\nu}$ the Palatini Ricci tensor, constructed in terms of an independent connection $\hat\Gamma^\alpha_{\mu\nu}$ as
\begin{equation}\label{eq:palatiniRab}
    \mathcal R_{\mu\nu}=\partial_\alpha\hat\Gamma^\alpha_{\mu\nu}-\partial_\nu\hat\Gamma^\alpha_{\mu\alpha}+\hat\Gamma^\alpha_{\alpha\beta}\hat \Gamma^\beta_{\mu\nu}-\hat\Gamma^\alpha_{\mu\beta}\hat\Gamma^\beta_{\alpha\nu},
\end{equation}
where $\partial_\mu\equiv \partial/\partial x^\mu$ denotes partial derivatives. Finally, $S_m\equiv \int_\Omega\sqrt{-g}\mathcal L_m d^4x$ is the action of the matter sector, where $\mathcal L_m$ represents the matter Lagrangian density which is minimally coupled to the gravitational sector.

\subsection{Scalar-tensor representation}

The action in Eq. \eqref{eq:gaction} can be conveniently rewritten in a dynamically-equivalent scalar-tensor representation through the addition of auxiliary scalar fields. The transformation to the scalar-tensor representation has been widely used in the literature, and proven useful in the simplification of the system of equations of motion. In the scalar-tensor representation, the action takes the form (see e.g. \cite{Rosa:2017jld} for the details on how to perform the transformation)
\begin{eqnarray}\label{eq:staction}
    S&=&\frac{1}{2\kappa^2}\int_\Omega\sqrt{-g}\left[\left(\varphi-\psi\right)R-\right.\nonumber \\
    &-&\left.\frac{3}{2\psi}\partial^\mu\psi\partial_\mu\psi-V\left(\varphi,\psi\right)\right]d^4x+S_m,
\end{eqnarray}
where the scalar fields $\varphi$ and $\psi$ and the interaction potential $V\left(\varphi,\psi\right)$ are defined as
\begin{equation}\label{eq:defscalar}
    \varphi=\frac{\partial f}{\partial R}, \qquad \psi=-\frac{\partial f}{\partial \mathcal R},
\end{equation}
\begin{equation}\label{eq:defpotential}
    V\left(\varphi,\psi\right)=-f\left(R,\mathcal R\right)+\varphi R-\psi \mathcal R
\end{equation}
We note that, for the transformation from the geometrical representation to the scalar-tensor representation to be well-defined, it is required that the determinant of the Hessian matrix of $f\left(R,\mathcal R\right)$ is non-zero. This is equivalent to requiring that Eqs. \eqref{eq:defscalar} are invertible. Indeed, these equations imply that the scalar fields $\varphi$ and $\psi$ can be written as functions of $R$ and $\mathcal R$ as $\varphi\left(R, \mathcal R\right)$ and $\psi\left(R,\mathcal R\right)$, and the transformation is well-defined if it is possible to write $R$ and $\mathcal R$ as functions of $\varphi$ and $\psi$ as $R\left(\varphi,\psi\right)$ and $\mathcal R\left(\varphi,\psi\right)$. This invertibility guarantees the uniqueness of the transformation (see Ref. \cite{Rosa:2017jld}).

The modified field equations in the scalar-tensor representation can be obtained through the variation of Eq. \eqref{eq:staction} with respect to the metric $g_{\mu\nu}$, which take the form (see e.g. \cite{Rosa:2018jwp})
\begin{eqnarray}
\label{eqn:fieldequations}
    \left(\varphi-\psi\right)G_{\mu\nu}&=&\kappa^2T_{\mu\nu}+\frac{3}{2\psi}\partial_\mu\psi\partial_\nu\psi-\frac{3}{4\psi}g_{\mu\nu}\partial^\alpha\psi\partial_\alpha\psi-\nonumber \\
    &-&\frac{1}{2}g_{\mu\nu}V+\left(\nabla_\mu \nabla_\nu-g_{\mu\nu}\Box\right)\left(\varphi-\psi\right),\label{eq:stfield}
\end{eqnarray}
where $G_{\mu\nu}=R_{\mu\nu}-\frac{1}{2}g_{\mu\nu}R$ is the Einstein's tensor, $\nabla_\mu$ represents the covariant derivatives, $\Box=\nabla_\mu\nabla^\mu$ is the d'Alembert operator, and $T_{\mu\nu}$ is the stress-energy tensor defined in terms of a variation of the matter Lagrangian as
\begin{equation}\label{eq:defTab}
    T_{\mu\nu}=-\frac{2}{\sqrt{-g}}\frac{\delta\left(\sqrt{-g}\mathcal L_m\right)}{\delta g^{\mu\nu}}.
\end{equation}
On the other hand, the equations of motion for the scalar fields $\varphi$ and $\psi$ can be obtained through a variation of Eq. \eqref{eq:staction} with respect to the scalar fields. These equations can be written in the form \cite{Rosa:2017jld}
\begin{equation}\label{eq:eomphi}
    \Box\varphi +\frac{1}{3}\left(2V-\psi V_\psi-\varphi V_\varphi\right)=\frac{8\pi T}{3},
\end{equation}
\begin{equation}\label{eq:eompsi}
    \Box\psi-\frac{1}{2\psi}\nabla_\mu\psi\nabla^\mu\psi-\frac{\psi}{3}\left(V_\varphi+V_\psi\right)=0,
\end{equation}
where we have introduced the notation $V_\varphi\equiv\partial V/\partial \varphi$ and $V_\psi\equiv\partial V/\partial\psi$, and $T=g^{\mu\nu}T_{\mu\nu}$ is the trace of the stress-energy tensor. Equations \eqref{eq:stfield}, \eqref{eq:eomphi}, and \eqref{eq:eompsi} constitute the system of equations of motion of the GHMPG in the scalar-tensor representation.

\section{Linearized system of equations}\label{sec:linear}

\subsection{Linear perturbations}

The linearized system of equations of motion can be obtained through the introduction of first-order perturbations in the metric $g_{\mu\nu}$ and scalar fields $\varphi$ and $\psi$. In what follows, $\epsilon\ll 1$ represents a perturbative parameter on which to perform a series expansion, and a bar $\bar{X}$ denotes the unperturbed value of a given quantity $X$. We assume that the background metric is Minkowski, i.e., $\bar g_{\mu\nu}=\eta_{\mu\nu}$, which implies that $\bar R = 0$, $\bar R_{\mu\nu}=0$, and consequently $\bar G_{\mu\nu}=0$. Under these assumptions, we write the metric $g_{\mu\nu}$ and the scalar fields $\varphi$ and $\psi$ as
\begin{eqnarray}\label{eq:perturbgs}
    g_{\mu\nu} &=& \eta_{\mu\nu}+\epsilon \delta g_{\mu\nu},\nonumber\\
    \varphi&=&\bar\varphi + \epsilon\delta\varphi,\\
    \psi&=&\bar\psi+\epsilon\delta\psi,\nonumber
\end{eqnarray}
where $\delta g_{\mu\nu}$ denotes the first-order perturbation of the metric, and $\delta\varphi$ and $\delta\psi$ denote the perturbations of the scalar fields, respectively. The Ricci tensor and the Ricci scalar take the forms
\begin{equation}\label{eq:perturbRab}
    R_{\mu\nu}=\epsilon \delta R_{\mu\nu}=\frac{1}{2}\epsilon\left(2\nabla^\alpha\nabla_{(\mu}\delta g_{\nu)\alpha}-\Box\delta g_{\mu\nu}-\nabla_\mu\nabla_\nu \delta g\right),
\end{equation}
\begin{equation}\label{eq:perturbR}
    R=\epsilon\delta R=\epsilon\left(\nabla_\mu\nabla_\nu\delta g^{\mu\nu}-\Box\delta g\right),
\end{equation}
where we introduced index symmetrization $2X_{(\mu\nu)}\equiv X_{\mu\nu}+X_{\nu\mu}$ and $\delta g=\eta_{\mu\nu}\delta g^{\mu\nu}$ is the trace of the metric perturbation. Consequently, the Einstein's tensor takes the form
\begin{equation}\label{eq:perturbG}
    G_{\mu\nu}=\epsilon\delta G_{\mu\nu}=\epsilon\left(\delta R_{\mu\nu}-\frac{1}{2}\eta_{\mu\nu}\delta R\right).
\end{equation}
The potential $V$ and its partial derivatives $V_\varphi$ and $V_\psi$ can be written to first-order in $\epsilon$ as
\begin{eqnarray}\label{eq:perturbV}
    V\left(\varphi,\psi\right)&=&\bar V+\epsilon\left(\bar V_{\varphi}\delta\varphi + \bar V_{\psi}\delta\psi\right),\nonumber\\
    V_\varphi\left(\varphi,\psi\right)&=&\bar V_\varphi+\epsilon\left(\bar V_{\varphi\varphi}\delta\varphi + \bar V_{\varphi\psi}\delta\psi\right),\\
    V_\psi\left(\varphi,\psi\right)&=&\bar V_\psi+\epsilon\left(\bar V_{\psi\varphi}\delta\varphi + \bar V_{\psi\psi}\delta\psi\right),\nonumber
\end{eqnarray}
where we have introduced the notation $\bar V\equiv V\left(\bar\varphi,\bar\psi\right)$ and similar for the partial derivatives of $V$, whereas the covariant derivatives and d'Alembert operators on the scalar fields take the forms
\begin{eqnarray}\label{eq:perturbds}
    \nabla_\mu\varphi&=&\epsilon\nabla_\mu\delta\varphi,\nonumber\\
    \nabla_\mu\psi&=&\epsilon\nabla_\mu\delta\psi,\\
    \Box\varphi&=&\epsilon\Box\delta\varphi,\nonumber \\
    \Box\psi&=&\epsilon \Box\delta\psi.\nonumber
\end{eqnarray}

Introducing the perturbations defined in Eqs. \eqref{eq:perturbgs} to \eqref{eq:perturbds} into the modified field equations in Eq. \eqref{eq:stfield} and the equations of motion for the scalar fields in Eqs. \eqref{eq:eomphi} and \eqref{eq:eompsi}, and keeping only the order-zero terms in $\epsilon$, one obtains the following three constraints to the background equations:
\begin{equation}\label{eq:background1}
    \eta_{\mu\nu}\bar V=16\pi \bar T_{\mu\nu},
\end{equation}
\begin{equation}\label{eq:background2}
    2\bar V-\bar\varphi\bar V_\varphi-\bar\psi\bar V_\psi=8\pi\bar T,
\end{equation}
\begin{equation}\label{eq:background3}
    \bar\psi\left(\bar V_\varphi+\bar V_\psi\right)=0.
\end{equation}
Equation Eq.\eqref{eq:background1} implies that, in vacuum, i.e., for $T_{\mu\nu}=0$, it is strictly necessary that $\bar V=0$. Consequently, one obtains through Eq.\eqref{eq:background2} that $\bar\varphi \bar V_\varphi + \bar\psi\bar V_\psi=0$. Finally, since $\bar\psi\neq 0$ to avoid divergences in the equations of motion, Eq. \eqref{eq:background3} implies that $\bar V_\varphi+\bar V_\psi=0$. This allows one to rewrite The last two equations as $\left(\bar\varphi-\bar\psi\right)\bar V_\varphi=0$. Since $\bar\varphi-\bar\psi\neq 0$ to preserve the well-definition of the scalar-tensor representation, we conclude that $\bar V_\varphi=0$ which implies that $\bar V_\psi=0$. Summarizing, the zero-order perturbation equations imply that the interaction potential must satisfy the following background constraints:
\begin{equation}\label{eq:constraints}
V\left(\bar\varphi,\bar\psi\right)=V_\varphi\left(\bar\varphi,\bar\psi\right)=V_\psi\left(\bar\varphi,\bar\psi\right)=0.
\end{equation}

Introducing the background constraints given in Eq. \eqref{eq:constraints} into the first-order perturbative equations, one obtains the following system of perturbative equations of motion
\begin{equation}\label{eq:fieldperturb}
    \delta G_{\mu\nu}=\left(\nabla_\mu\nabla_\nu-\eta_{\mu\nu}\Box\right)\left(\frac{\delta\varphi-\delta\psi}{\bar\varphi-\bar\psi}\right),
\end{equation}
\begin{equation}\label{eq:eomphiperturb}
    \left(\Box-M_\varphi^2\right)\delta\varphi=\Lambda_\varphi\delta\psi,
\end{equation}
\begin{equation}\label{eq:eompsiperturb}
    \left(\Box-M_\psi^2\right)\delta\psi=\Lambda_\psi\delta\varphi,
\end{equation}
where we have defined the following quantities for convenience
\begin{eqnarray}\label{eq:masscoupling}
    M_\varphi^2&=&\frac{1}{3}\left(\bar\varphi\bar V_{\varphi\varphi}+\bar\psi\bar V_{\varphi\psi}\right),\nonumber\\
    M_\psi^2&=&\frac{1}{3}\bar\psi\left(\bar V_{\varphi\psi}+\bar V_{\psi\psi}\right),\\
    \Lambda_\varphi&=&\frac{1}{3}\left(\bar\varphi \bar V_{\psi\varphi}+\bar\psi \bar V_{\psi\psi}\right),\nonumber\\
    \Lambda_\psi&=&\frac{1}{3}\bar\psi\left(\bar V_{\varphi\varphi}+\bar V_{\psi\varphi}\right).\nonumber
\end{eqnarray}
where $M_\varphi$ and $M_\psi$ play the role of the masses of the perturbations $\delta\varphi$ and $\delta\psi$, respectively, and $\Lambda_\varphi$ and $\Lambda_\psi$ play the role of the source terms. It is interesting to note that if we chose the potential $V\left(\varphi,\psi\right)$ in such a way that $\bar V_{\varphi\varphi}=\bar V_{\psi\psi}=\bar V_{\varphi\psi}=0$, one obtains $\Box\delta\varphi=\Box\delta\psi=0$, and consequently $\Box R_{\mu\nu}=\Box R=0$, i.e., one recovers the propagation of gravitational waves in vacuum for GR. Such a choice of potential, e.g. featuring only cubic terms (or higher) in the scalar fields, thus renders the theory unfalsifiable from the point of view of gravitational wave propagation in vacuum. Any deviation from GR requires the presence of quadratic terms in the potential $V$, i.e., the potential should be written in the form
\begin{eqnarray}\label{eq:quad_potential}
    &&V\left(\varphi, \psi\right)=c_1\epsilon^2 \delta\varphi^2 + c_2 \epsilon^2 \delta\psi^2+c_3\epsilon^2\delta\varphi\delta\psi\\
    &&= c_1\left(\varphi-\bar\varphi\right)^2+c_2\left(\psi-\bar\psi\right)^2+c_3\left(\varphi-\bar\varphi\right)\left(\psi-\bar\psi\right) \nonumber,
\end{eqnarray}
for some arbitrary constants $c_i\neq 0$. Further on, we impose additional constraints on this potential based on the equations of motion for the scalar modes.

\subsection{Tensor modes}

To analyze the tensor modes of gravitational wave propagation, it is necessary to select an appropriate gauge to set. For this purpose, we define an auxiliary quantity $\chi_{\mu\nu}$ defined as
\begin{equation}\label{eq:defchi}
    \chi_{\mu\nu}\equiv \delta g_{\mu\nu}-\frac{1}{2}\eta_{\mu\nu}\delta g-\eta_{\mu\nu}\left(\frac{\delta\varphi-\delta\psi}{\bar\varphi-\bar\psi}\right).
\end{equation}
Introducing the definition of $\chi_{\mu\nu}$ into Eq. \eqref{eq:fieldperturb}, one obtains
\begin{equation}\label{eq:auxgauge1}
    2\nabla^\alpha\nabla_{(\mu}\chi_{\nu)\alpha}-\eta_{\mu\nu}\nabla_{\alpha}\nabla_\beta\chi^{\alpha\beta}-\Box\chi_{\mu\nu}=0.
\end{equation}
Equation \eqref{eq:auxgauge1} indicates that the appropriate choice of gauge is $\nabla^\mu\chi_{\mu\nu}=0$, which reduces the equation to $\Box\chi_{\mu\nu}=0$. Furthermore, taking the trace of Eq. \eqref{eq:fieldperturb} and using Eq. \eqref{eq:perturbR}, one obtains
\begin{equation}\label{eq:auxgauge2}
    \Box\left(\frac{\delta\varphi-\delta\psi}{\bar\varphi-\bar\psi}\right)=-\frac{1}{4}\Box\delta g.
\end{equation}
Combining Eqs. \eqref{eq:auxgauge1} and \eqref{eq:auxgauge2} under the choice of gauge $\nabla^\mu\chi_{\mu\nu}=0$ leads to the simplified propagation equation for the metric perturbation as
\begin{equation}\label{eq:propmetric}
    \Box\chi_{\mu\nu}=\Box\left(\delta g_{\mu\nu}-\frac{1}{4}\eta_{\mu\nu}\delta g\right)=0.
\end{equation}

The solution of the equation above can be written in terms of plane waves as \cite{Gong:2017bru}
\begin{equation}
\chi_{\mu\nu}=e^{\pm}_{\mu\nu}\exp\left(\pm i k_\alpha x^\alpha\right),
\label{chi_munu}
\end{equation}
where $e^{\pm}_{\mu\nu}$ represent the amplitudes of the outgoing ($+$) and ingoing ($-$) waves, and $k^\mu$ is the wave vector. The wave vector and the amplitudes satisfy the orthogonality conditions 
\begin{eqnarray}\label{eq:normstensor}
    &&k^\mu e_{\mu\nu}=0, \nonumber  \\
    &&\eta^{\mu\nu}e_{\mu\nu}=0, \nonumber\\
    &&\eta^{\mu\nu}k_\mu k_\nu=0.
\end{eqnarray}
One may choose, without loss of generality, a preferred direction of propagation $k_\mu=\left(-\omega, k, 0, 0\right)$, where $\omega$ is the angular frequency of the wave and $k$ is the corresponding wave number. The orthogonality conditions in Eq. \eqref{eq:normstensor} imply that $\omega^2=k^2$, which implies that $v=\frac{d\omega}{d k}=\pm 1$, i.e., the tensor modes propagate at the speed of light. Under these considerations, the auxiliary tensor $\chi_{\mu\nu}$ and the metric perturbation $\delta g_{\mu\nu}$ take the forms
\begin{equation}
    \chi_{\mu\nu}=\begin{pmatrix}
        0 & 0 & 0 & 0 \\
        0 & 0 & 0 & 0 \\
        0 & 0 & \frac{1}{2}\left(\delta g_{yy}-\delta g_{zz}\right) & \delta g_{yz} \\
        0 & 0 & \delta g_{yz} & \frac{1}{2}\left(\delta g_{zz}-\delta g_{yy}\right)
    \end{pmatrix},
\end{equation}
\begin{equation}\label{eq:gabprop}
    \delta g_{\mu\nu}=\begin{pmatrix}
        -\frac{1}{2}\left(\delta g_{yy}+\delta g_{zz}\right) & 0 & 0 & 0 \\
        0 & \frac{1}{2}\left(\delta g_{yy}+\delta g_{zz}\right) & 0 & 0 \\
        0 & 0 & \delta g_{yy} & \delta g_{yz} \\
        0 & 0 & \delta g_{yz} & \delta g_{zz} 
    \end{pmatrix}.
\end{equation}
The metric perturbation in Eq. \eqref{eq:gabprop} presents three degrees of freedom associated with the metric perturbation components $\delta g_{yy}$, $\delta g_{zz}$, and $\delta g_{yz}$. The metric perturbation component $\delta g_{yz}$ characterizes the usual $\times$-polarization also present in GR. However, the components $\delta g_{yy}$ and $\delta g_{zz}$ cease to be symmetric, and thus they contribute with both a transversal and a longitudinal polarizations. At this point, one can verify that the GHMPG encompasses GR as a particular solution. Indeed, when one uses the usual gauge condition in GR, $\nabla^{\mu} h_{\mu\nu} \equiv \nabla^{\mu} \left(\delta g_{\mu\nu} - \frac{1}{2} \eta_{\mu\nu} \delta g\right) = 0$ (which is equivalent to $\nabla ^{\mu} \chi_{\mu\nu} = 0$ in our case), the analogous of Eq. \eqref{eq:propmetric} has a factor of $\frac{1}{2}$ instead of $\frac{1}{4}$ (the latter is the result of the uncoupling of the extra scalar degree of freedom of GHMPG from the tensor modes, a characteristic that can arise in modified gravity theories \cite{Gomes:2018}). Therefore, in such a gauge, the metric perturbation in GR satisfies the condition $\delta g=0$ (which together with the previous transverse condition is the well-known traceless-transverse gauge, or TT-gauge). Then, if one imposes this condition on the metric perturbation given in Eq. \eqref{eq:gabprop}, $\delta g_{tt}$ and $\delta g_{xx}$ cancel out, forcing $\delta g_{yy}=-\delta g_{zz}$. The additional longitudinal polarization mode is thus removed, and the additional transversal polarization mode reduces to the usual $+$-polarization in GR. This clarifies that the GR solution exists as a particular case of the general GHMPG solution.
\subsection{Scalar modes}

To analyze the two scalar modes present Eqs. \eqref{eq:eomphiperturb} and \eqref{eq:eompsiperturb}, it is convenient to follow a diagonalization procedure in order to decouple these modes. This system can be written in a matrix form as
\begin{equation}\label{eq:matrixsys1}
    \left(I_{2\times2}\Box-\mathcal M\right)\Theta=0,
\end{equation}
where $I_{2\times2}$ is the two-dimensional identity matrix, and the matrices $\mathcal M$ and $\Theta$ are defined as
\begin{equation}
    \mathcal M = \begin{pmatrix}
    M_\varphi^2 & \Lambda_\varphi \\
    \Lambda_\psi & M_\psi^2
    \end{pmatrix},\qquad \Theta=\begin{pmatrix}
        \delta\varphi \\
        \delta\psi \end{pmatrix}.
\end{equation}
Let $\mathcal S$ be the matrix of the eigenvectors of $\mathcal M$. Furthermore, define $\mathcal M_D=\mathcal S^{-1} \mathcal M \mathcal S$ and $\Theta_D=\mathcal S^{-1}\Theta$. Under these assumptions, Eq.\eqref{eq:matrixsys1} can be rewritten in the diagonalized form
\begin{equation}\label{eq:matrixsys2}
    \left(I_{2\times2}\Box-\mathcal M_D\right)\Theta_D=0,
\end{equation}
where the matrices $\mathcal M_D$ and $\Theta_d$ are given by
\begin{equation}
    \mathcal M_D = \begin{pmatrix}
        M_\Phi^2 & 0 \\
        0 & M_\Psi^2
    \end{pmatrix},\qquad \Theta_D=\begin{pmatrix}
        \delta\Phi \\
        \delta\Psi
    \end{pmatrix},
\end{equation}
and where we have introduced the following notation for the redefined masses $M_\Phi^2$ and $M_\Psi^2$,
\begin{eqnarray}
    M_\Phi^2&=&\Lambda_\psi M_-,\nonumber \\
    M_\Psi^2&=&\Lambda_\psi M_+,\nonumber\\
    M_\pm&=&M_0\pm M_1, \\
    M_0&=&\frac{M_\varphi^2-M_\psi^2}{2\Lambda_\psi},\nonumber\\
M_1&=&\sqrt{M_0^2+\frac{\Lambda_\varphi}{\Lambda_\psi}}\nonumber
\label{Masses}
\end{eqnarray}
and the redefined scalar field perturbations $\delta\Phi$ and $\delta\Psi$,
\begin{eqnarray}
    \delta\Phi&=&-\frac{1}{2M_1}\left(\delta\varphi-M_+\delta\psi\right),\nonumber \\
    \delta\Psi&=&-\frac{1}{2M_1}\left(\delta\varphi-M_-\delta\psi\right).
    \label{deltas}
\end{eqnarray}
Through the method outlined above, the coupled system of Eqs. \eqref{eq:eomphiperturb} and \eqref{eq:eompsiperturb} can be conveniently rewritten in the form of an uncoupled Klein-Gordon system of the form
\begin{eqnarray}\label{eq:propscalar}
    \left(\Box-M_\Phi^2\right)\delta\Phi=0,\nonumber \\
    \left(\Box-M_\Psi^2\right)\delta\Psi=0.
\end{eqnarray}

Similarly to the tensor modes, the solutions of the two equations above can be written in terms of plane waves as\cite{Gong:2017bru}
\begin{eqnarray}
\delta\Phi&=&\Phi^\pm\exp\left(\pm i p_\mu x^\mu\right), \nonumber \\
\delta\Psi&=&\Psi^\pm\exp\left(\pm i q_\mu x^\mu\right),
\label{kleingordonsols}
\end{eqnarray}
where $\Phi^\pm$ and $\Psi^\pm$ represent the amplitudes of the outgoing ($+$) and ingoing ($-$) waves, and $p^\mu$ and $q^\mu$ are the wave vectors. Due to the fact that the scalar modes are massive, the normalization conditions for the wave vectors in this case become 
\begin{eqnarray}\label{eq:normscalar}
    \eta^{\mu\nu}p_\mu p_\nu =-M_\Phi^2, \nonumber \\
    \eta^{\mu\nu}q_\mu p_\nu =-M_\Psi^2.
\end{eqnarray}
Choosing, without loss of generality, a preferred direction of propagation $p^\mu=\left(-\omega_\Phi, k_\Phi, 0, 0\right)$ and $q^\mu=\left(-\omega_\Psi, k_\Psi, 0, 0\right)$, the normalization conditions in Eq. \eqref{eq:normscalar} imply that the propagation velocities are
\begin{eqnarray}
    v_\Phi=\frac{\partial \omega_\Phi}{\partial k_\Phi}=\pm \frac{k_\Phi}{\sqrt{k_\Phi^2+M_\Phi^2}},\nonumber \\
    v_\Psi=\frac{\partial \omega_\Psi}{\partial k_\Psi}=\pm \frac{k_\Psi}{\sqrt{k_\Psi^2+M_\Psi^2}}.\nonumber
\end{eqnarray}

One can thus observe that, since $M_\Phi^2\geq0$ and $M_\Psi^2\geq0$, the propagation velocities are bounded by $|v_\Phi|\leq 1$ and $|v_\Psi|\leq 1$, with the equal sign being only applicable for the particular choices of the potential $V\left(\phi,\psi\right)$ for which $M_\Phi$ and $M_\Psi$ vanish. Following the definitions in Eqs. \eqref{eq:masscoupling} and \eqref{Masses} and the potential obtained in Eq. \eqref{eq:quad_potential}, one verifies that the masses $M_\Phi^2$ and $M_\Psi^2$ can be written in the form
\begin{eqnarray}
    M_\Phi^2=\frac{1}{3} \left(A - \sqrt{B}\right), \\
    M_\Psi^2=\frac{1}{3}\left(A + \sqrt{B}\right), \nonumber
\end{eqnarray}
where the constants $A$ and $B$ can be written in terms of the constants $c_i$ and the background scalar fields $\bar\varphi$ and $\bar\psi$ as
\begin{eqnarray}
    A&=&c_1\bar\varphi-c_2\bar\psi,\\
    B&=&b_1 \bar\varphi^2+b_2\bar\psi^2+b_3 \bar\varphi\bar\psi, \nonumber
\end{eqnarray}
where we have defined the coefficients $b_i$ as
\begin{eqnarray}
    b_1 &=& c_1^2, \nonumber \\
    b_2 &=& c_2\left(4c_1+c_2+2c_3\right), \\
    b_3 &=& c_3^2+2c_1\left(c_3-c_2\right). \nonumber
\end{eqnarray}
The masses $M_\Phi^2$ and $M_\Psi^2$ are simultaneously zero whenever the constants $A$ and $B$ satisfy $A=0$ and $B=0$. Solving $A=0$ with respect to $c_2$, replacing the result into $B=0$ and solving for $c_3$, one obtains the following constraints for the constants $c_2$ and $c_3$:
\begin{equation}\label{eq:constants_sol}
    c_2 = c_1\frac{\bar\varphi}{\bar\psi}, \qquad
    c_3 = -2 c_1.
\end{equation}
Introducing the constants given in Eq.\eqref{eq:constants_sol} into the potential given in Eq. \eqref{eq:quad_potential}, one obtains the following expression for the potential
\begin{eqnarray}\label{eq:potential_fine}
&&V\left(\varphi,\psi\right)=c_1\epsilon^2\left(\delta\varphi^2+\frac{\bar\varphi}{\bar\psi}\delta\psi^2-2\delta\varphi\delta\psi\right) \\
&&=c_1\left[\left(\varphi-\bar\varphi\right)^2+\frac{\bar\varphi}{\bar\psi}\left(\psi-\bar\psi\right)^2-2\left(\varphi-\bar\varphi\right)\left(\psi-\bar\psi\right)\right].\nonumber
\end{eqnarray}
This result implies that if one chooses a potential of the form given in Eq.\eqref{eq:potential_fine}, the masses of the scalar modes vanish and the scalar modes propagate at the speed of light.

\section{Polarization States}\label{sec:newman}

\subsection{The Newman-Penrose formalism}

In addition to the linearization formalism for analyzing the polarization states of a gravitational wave far from the source, Newman and Penrose developed an alternative formalism based on tetrads as the building blocks of gravity. From the decomposition of the Riemann tensor $R_{\rho\sigma\mu\nu}$ into its irreducible parts
\begin{eqnarray}
R_{\rho\sigma\mu\nu}&=& C_{\rho\sigma\mu\nu}+\frac{1}{2}\left(g_{\sigma\nu}R_{\rho\mu}+g_{\rho\mu}R_{\sigma\nu}-g_{\sigma\mu}R_{\rho\nu}-\right.\nonumber\\
&&\left.g_{\rho\nu}R_{\sigma\mu}\right)+\frac{R}{6}\left(g_{\rho\nu}g_{\sigma\mu}-g_{\rho\mu}g_{\sigma\nu}\right) ~,
\end{eqnarray}
where $C_{\rho\sigma\mu\nu}$ is the Weyl tensor, one can write the so-called Newman-Penrose (NP) quantities, which are based on the ten independent components of the Weyl tensor ($\Psi_i$), the nine traceless Ricci scalar ($\Phi_i$) and the curvature scalar ($\Lambda$). From this set of quantities, one can find the linearly independent ones for a given gravity theory and characterize the spacetime. Furthermore, Newman and Penrose chose a basis of null tetrads on which the previous quantities can be written, namely
\begin{eqnarray}
    &&n_{\mu}=\frac{1}{\sqrt{2}}(1,1,0,0) ~ , \quad l_{\mu}=\frac{1}{\sqrt{2}}(1,-1,0,0) ~ ,\nonumber\\
    &&m_{\mu}=\frac{1}{\sqrt{2}}(0,0,1,i) ~ , \quad \bar{m}_{\mu}=\frac{1}{\sqrt{2}}(0,0,1,-i) ~ ,
\end{eqnarray}
where the null tetrads satisfy the following orthogonality conditions
\begin{equation}
    n_{\alpha} l^{\alpha} = - m_{\alpha} \bar{m}^{\alpha}= - 1~,
\end{equation}
and the scalar product between any other pair of tetrads aside from these vanishes. In this basis, and using the sign convention of \cite{book:Frolov,Polarisations}, the NP quantities can be written as
\begin{eqnarray}
    &&\Psi_0 = C_{nmnm} ~, \quad \Psi_1 = C_{nlnm}~,\nonumber\\
    &&\Psi_2 = C_{nm\bar{m}l}~, \quad \Psi_3 = C_{nl\bar{m}l} ~,\nonumber\\
    &&\Psi_4 = C_{l\bar{m}l\bar{m}}~, \nonumber\\
    &&\Phi_{00} = \frac{1}{2}R_{nn} ~, \quad \Phi_{11} = \frac{1}{4}\left(R_{nl}+R_{m\bar{m}}\right)~,\label{eq:NPdefinitions}\\
    &&\Phi_{22} = \frac{1}{2}R_{ll} ~, \quad \Phi_{01}=\Phi_{10}^*=\frac{1}{2}R_{nm} ~,\nonumber\\
    &&\Phi_{02}=\Phi_{20}^* =\frac{1}{2}R_{mm} ~, \quad \Phi_{12}=\Phi_{21}^* = \frac{1}{2}R_{lm} ~,\nonumber\\
    &&\Lambda=\frac{R}{24}\nonumber ~,
\end{eqnarray}
where the quantities written in the NP basis mean that any tensor is expressed as $A_{abc...}=A_{\mu\nu\rho \dots}a^{\mu}b^{\nu}c^{\rho}\dots$, where $a,b,c,\dots$ are vectors of the null-complex tetrad basis $(n,l,m,\bar{m})$ and $\mu,\nu,\rho,\dots$ run over spacetime indices.

Eardley, Lee, and Lightman have developed a formalism for classifying the gravitational wave solution for metric theories allowing for plane null gravitational wave solutions, that is, with massless modes, in which the NP quantities reduce to only six real independent components, namely the set $\{\Psi_2,~\Psi_3,~\Psi_4,~\Phi_{22}\}$ (the other nonvanishing components can be shown to be  linear combinations of the elements of this set). The associated symmetry group is a subgroup of the Lorentz group known as the “little group” $E(2)$ that leaves the wave-vector invariant. Under the action of the rotation group, the values of the helicity for each of the components of the set are, respectively, $\{0,~\pm 1, \pm 2, 0\}$. Hence, $\Psi_2$ represents a longitudinal mode, $\Psi_3$ can be split into the real and purely imaginary parts responsible for the vectorial $y-$ and $z-$ modes, $\Psi_4$ can also be split as the former leading to the tensor $+$ and $\times$ polarizations, and $\Phi_{22}$ represents the scalar breathing mode \cite{ELL1,ELL2}.

For gravity theories with massive gravitational wave polarization modes, the formalism outlined above cannot be straightforwardly applied. As we have found in Sec. \ref{sec:linear} through the linearization method, the GHMPG features two massive scalar modes that can be decoupled. Thus, we shall use the method of inverting the NP quantities as functions of the polarization states as done in Ref. \cite{Polarisations}.

\subsection{Massive Propagating Modes}
For metric theories with massive propagating gravitational wave modes, the NP formalism must be reformulated. We start by inspecting the only measurable quantity in gravitational wave detection, namely the so-called Riemann field, obtained from the electric components of the Riemann tensor \cite{Polarisations}. The linearized Riemann tensor reads
\begin{equation}
    \delta R^{\alpha}_{\ \mu\beta\nu}=\frac{1}{2}\eta^{\alpha\sigma}\left(\delta g_{\sigma\nu , \mu\beta}-\delta g_{\mu\nu,\sigma\beta}+\delta g_{\sigma\beta , \mu\nu}-\delta g_{\mu\beta , \sigma\nu}\right)~,
\end{equation}
from which the electric components are
\begin{equation}
    \delta R_{itjt}=\frac{1}{2}\left(\delta g_{it,jt}-\delta g_{jt,it}+\delta g_{ij,tt} - \delta g_{tt,ij}\right)~.
\end{equation}
On the other hand, the geodesic deviation equation is
\begin{equation}
    \ddot{x}^i=-\delta R^i_{\ tjt}x^j ~.
\end{equation}

Choosing, without loss of generalization, the propagation of the gravitational wave along the $x-$ axis, the Riemann field is a function of the retarded time $t_r=t-x/v$, with $v$ the propagation speed of the gravitational wave. We introduce the driving-force matrix \cite{ELL1,Polarisations}
\begin{equation}
S_{ij}\left(t_r\right)=R_{titj}\left(t_r\right) ~,
\end{equation}
and the six basis polarization matrices as
\begin{eqnarray}
  &&E_1 = \begin{pmatrix}
        1 & 0 & 0 \\
        0 & 0 & 0 \\
        0 & 0 & 0
    \end{pmatrix},\qquad E_2=\begin{pmatrix}
        0 & 1 & 0 \\
        1 & 0 & 0 \\
        0 & 0 & 0
    \end{pmatrix},\nonumber\\
    &&E_3 = \begin{pmatrix}
        0 & 0 & 1 \\
        0 & 0 & 0 \\
        1 & 0 & 0
    \end{pmatrix},\qquad E_4=\frac{1}{2}\begin{pmatrix}
        0 & 0 & 0 \\
        0 & 1 & 0 \\
        0 & 0 & -1
    \end{pmatrix},\\
    &&E_5 = \begin{pmatrix}
        0 & 0 & 0 \\
        0 & 0 & 1 \\
        0 & 1 & 0
    \end{pmatrix},\qquad E_6=\frac{1}{2}\begin{pmatrix}
        0 & 0 & 0 \\
        0 & 1 & 0 \\
        0 & 0 & 1
    \end{pmatrix}.\nonumber
\end{eqnarray}

In the basis of the polarization matrices, the driving-force takes the form
\begin{eqnarray}
S\left(t_r\right)&&=\sum_{A=1}^6 p_A\left(t_r,e'_x\right)E_A\left(e'_x\right) =\nonumber\\
    &&=\begin{pmatrix}
        p_1 & p_2 & p_3 \\
        p_2 & \frac{1}{2}(p_4+p_6) & p_5 \\
        p_3 & p_5 & \frac{1}{2}(-p_4+p_6)
    \end{pmatrix} ~,
\end{eqnarray}
where $p_1, \dots , p_6$ are the longitudinal, vectorial y-, vectorial z-, $+$, $\times$, and breathing polarization modes, respectively.

Thus, for plane gravitational waves, one can cast the polarization modes as functions of the NP quantities as follows
\begin{eqnarray}
    p_1^{(l)} && = 6\Psi_2-\left[2(\Psi_2+2\Lambda)\right]~,\nonumber\\
    p_2^{(y)}&& = 2 \text{Re}(\Psi_3)-\left[2\text{Re}(\Psi_1)\right] ~,\nonumber\\
    p_3^{(z)}&& = - 2 \text{Im}(\Psi_3) - [2 \text{Im}(\Psi_1)] ~,\nonumber\\
    p_4^{(+)}&& = \text{Re}(\Psi_4) + \left[\text{Re}(\Psi_0) - 2\text{Re}(\Phi_{02})\right] ~,\label{eq:polarizationstates}\\
    p_5^{(\times)}&& = -\frac{1}{2}\text{Im}(\Psi_4) + \left[\frac{1}{2}\text{Im}(\Psi_0) - \text{Im}(\Phi_{02})\right] ~,\nonumber\\
    p_6^{(b)}&& = \Phi_{22} - \left[2(\Psi_2 + 2\Lambda) - \Phi_{00}\right] ~,\nonumber
\end{eqnarray}
where the quantities inside square brackets vanish for the massless case. 

One can now plug the definitions of the NP-quantities in Eqs. \eqref{eq:NPdefinitions} into the polarization states given in Eqs. \eqref{eq:polarizationstates}. As for plane gravitational wave solutions, one can write the polarization states as functions of the components of the Einstein tensor and components of the Riemann tensor as \cite{Polarisations}
\begin{eqnarray}
    p_1^{(l)} && = \frac{1}{2}\left(G_{tt}-G_{xx}+G_{yy}+G_{zz}\right)~,\nonumber\\
    p_2^{(y)}&& = -G_{xy} ~,\nonumber\\
    p_3^{(z)}&& = - G_{xz} ~,\nonumber\\
    p_4^{(+)}&& = -\left(G_{yy}-G_{zz}\right)+R_{xyxy}-R_{xzxz} ~,\\
    p_5^{(\times)}&& = -G_{yz}+R_{xyxz} ~,\nonumber\\
    p_6^{(b)}&& = G_{xx} ~,\nonumber
\end{eqnarray}

Therefore, using the field equations in Eq. (\ref{eqn:fieldequations}) with $T_{\mu\nu}=0$ to expand the terms proportional to the Einstein's tensor, and noting that the derivatives of the scalar fields with respect to $y$ and $z$ vanish, we obtain:
\begin{eqnarray}
    p_1^{(l)} && = \frac{3}{4\psi(\varphi-\psi)}\left[(\partial_t\psi)^2-(\partial_x\psi)^2-\varrho\Xi+\right.\nonumber\\
    &&\left.+(\partial_t^2+\partial_x^2)(\varphi-\psi)\right]~,\nonumber\\
    p_2^{(y)}&& = \frac{3g_{xy}}{2\psi(\varphi-\psi)}\Xi ~,\nonumber\\
    p_3^{(z)}&& = \frac{3g_{xz}}{2\psi(\varphi-\psi)}\Xi ~,\nonumber\\
    p_4^{(+)}&& = \frac{3(g_{yy}-g_{zz})}{2\psi(\varphi-\psi)}\Xi+R_{xyxy}-R_{xzxz} ~,\label{eq:polarsimpler}\\
    p_5^{(\times)}&& = \frac{3g_{yz}}{2\psi(\varphi-\psi)}\Xi+R_{xyxz} ~,\nonumber\\
    p_6^{(b)}&& = \frac{3}{2\psi(\varphi-\psi)}\left[(\partial_x\psi)^2+\partial_x^2(\varphi-\psi)-g_{xx}\Xi\right]~,\nonumber
\end{eqnarray}
where we have introduced the following auxiliary quantities:
\begin{equation}
    \varrho\equiv g_{tt}-g_{xx}+g_{yy}+g_{zz}\nonumber,
\end{equation}
\begin{equation}
    \Xi\equiv \frac{1}{2}\partial^{\alpha}\psi\partial_{\alpha}\psi+\frac{\psi}{3}V+\frac{2\psi}{3}\square(\varphi-\psi)\nonumber.
\end{equation}

Equations \eqref{eq:polarsimpler} represent the most general description of gravitational wave polarization modes in the GHMPG, since the assumption of propagation in the $x$ direction is arbitrary. To proceed with the analysis of these modes, it is necessary to break generality either imposing explicit forms for the quantities appearing in the equations, e.g. choosing a particular form of the theory, or to impose additional constraints on these quantities through a comparison with observational data.

\subsection{Connection with the linearization method}

To finalize, let us perform a bridge between the results obtained through the analysis of the linearized system of equations and the analysis through the Newman-Penrose formalism. For this purpose, consider that one is far from the source, in which the linearization method around a Minkowski background is an adequate approximation of the fully non-linear analysis. In this limit, the polarization modes are given in terms of the components of the perturbed Riemann tensor. Hence, for the metric solution in Eq. \eqref{eq:gabprop}, one obtains
\begin{eqnarray}
    p_1^{(l)} && = -\frac{1}{4} \left(\delta g_{yy,tt}+\delta g_{zz,tt}-\delta g_{yy,xx}-\delta g_{zz,xx}\right)~,\nonumber\\
    p_2^{(y)}&& = 0 ~,\nonumber\\
    p_3^{(z)}&& = 0 ~,\nonumber\\
    p_4^{(+)}&& = -\frac{1}{2} \left(\delta g_{yy,tt}-\delta g_{zz,tt}\right)~,\\
    p_5^{(\times)}&& = -\frac{1}{2} \delta g_{yz,tt}~,\nonumber\\
    p_6^{(b)}&& = -\frac{1}{2} \left(\delta g_{yy,tt}+\delta g_{zz,tt}\right)~,\nonumber
    \label{polarizations_lin}
\end{eqnarray}
These results confirm that we have four polarization states but only three degrees of freedom, associated with $(\delta g_{yy}+ \delta g_{zz})$, $(\delta g_{yy}-\delta g_{zz})$ and $\delta g_{yz}$, which is in agreement with the results of Sec. \ref{sec:linear}. Moreover, these results agree with the method developed in Ref. \cite{Dong:2025ddi}.

\section{Constraints and implications}
\label{sec:implications}

For the particular case of plane gravitational waves propagating along the $x$ direction, which we consider to be an acceptable approximation given the large distance to the sources of the gravitational wave signals detected, the displacement is constant along each plane orthogonal to the propagation direction. Hence we expect that $g_{xy}=0=g_{xz}$, a result that is in agreement with the linearized case (see Eq. \eqref{eq:gabprop}), and therefore no vector polarization modes arise in the full non-linear regime of the theory, i.e. $p_2^{(y)}=p_3^{(z)}=0$. 

Additional simplifications to the results presented in Eqs. \eqref{eq:polarsimpler} can be obtained by taking into account the constrains on the scalar fields obtained from a cosmological analysis of the theory. Indeed, this gravity model has been thoroughly studied in a cosmological context through the dynamical systems approach \cite{Rosa:2024pzo}, and recent results show that the fields $\varphi$ and $\psi$ are exponentially suppressed during the matter-dominated epoch, reaching approximately constant values at late times, i.e. $\varphi,\psi \approx \text{const}$. Furthermore, the solar system dynamics constrain the scalar fields to $\varphi-\psi\approx 1$. As such, the potential $V$ is responsible for late-time cosmic acceleration and plays the role of a cosmological constant at late times, whose value is constrained to $V\approx 10^{-122}/m_P^2$, with $m_P$ the Planck mass $m_P=\sqrt{\hbar G/c^3}$. 

Taking the cosmological results into account at non-linear level, we expect that at present times, since $\partial_\mu\varphi\approx 0$, $\partial_\mu\psi\approx 0$, and $V\sim 10^{-122}$, and thus $\Xi\approx 0$, the theory may exhibit highly-suppressed scalar modes, $p_1^{(l)}$, $p_6^{(b)}$, and tensor polarizations which are essentially dominated by the Riemann tensor components. Even so, it is remarkable that the $+$ polarization is not necessarily identical to the one in GR, as it is not required that $R_{xyxy}=-R_{xzxz}$, and hence we there is a possible "smoking gun" for distinguishing between GHMPG and GR. However, it is important to note that since GR exists as a particular solution of GHMPG, in the absence of potentially measurable differences between $R_{xyxy}$ and $-R_{xzxz}$, this analysis renders the GHMPG unfalsifiable from the point of view of gravitational wave propagation in the plane gravitational wave approximation. 

Indeed, the best one can do is to estimate the amplitude of the polarization states present in Eq. \eqref{polarizations_lin} up to the maximum frequency of each present and future detector. To do so, one must write each mode in terms of the angular frequency of the wave $\omega$. For the $p_5^{(\times)}$ mode, since $\delta g_{yz}=\chi_{yz} \Rightarrow \delta g_{yz,tt} = \chi_{yz,tt}$, one needs to differentiate $\chi_{yz}$ with respect to the time coordinate twice using Eq. \eqref{chi_munu} in order to find $\delta g_{yz,tt}$. In addition, we simplify $p_4^{(+)}$ by using both $\chi_{yy}$ and $\chi_{zz}$, differentiate them twice with respect to time, and then manipulate the resulting system of equations, which directly yields $\delta g_{yy,tt}-\delta g_{zz,tt}$. However, it is not possible to estimate in a general manner the breathing mode $p_6^{(b)}$, as we do not possess enough information to simplify $\delta g_{yy,tt}+\delta g_{zz,tt}$. Finally, for the longitudinal mode $p_1^{(l)}$, by using Eq. \eqref{eq:auxgauge2} while considering that the solar system constraint remains valid in the Minkowski background $\bar\varphi-\bar\psi \approx 1$ and that the derivatives $\delta g_{ii,jj}$, with $i=y,z$ and $j=y,z$ can be neglected far from the source, one obtains the following relation
\begin{eqnarray}
\frac{1}{2} \left(\delta g_{yy,tt}+\delta g_{zz,tt}-\delta g_{yy,xx}-\delta g_{zz,xx}\right) && \nonumber\\
= \delta \varphi_{,xx} - \delta \varphi_{,tt} - \left(\delta \psi_{,xx} -\delta \psi_{,tt} \right).
\end{eqnarray}
Hence, the polarization modes in the linear limit are given by
\begin{eqnarray}
    p_1^{(l)} && = -\frac{1}{2} \left[\delta \varphi_{,xx} - \delta \varphi_{,tt} - \left(\delta \psi_{,xx} -\delta \psi_{,tt} \right)\right]~,\nonumber\\
    p_2^{(y)}&& = 0 ~,\nonumber\\
    p_3^{(z)}&& = 0 ~,\nonumber\\
    p_4^{(+)}&& = \pm\frac{1}{2} \omega^2 \left(\delta g_{yy}-\delta g_{zz}\right)~,\\
    p_5^{(\times)}&& = \pm \frac{1}{2} \omega^2 \delta g_{yz}~,\nonumber\\
    p_6^{(b)}&& = -\frac{1}{2} \left(\delta g_{yy,tt}+\delta g_{zz,tt}\right)~.\nonumber
\end{eqnarray}
In fact, one could fully estimate the longitudinal mode in the linear limit by using Eqs. \eqref{deltas} and \eqref{kleingordonsols} to obtain
\begin{eqnarray}
\delta\psi=\frac{2M_1}{M_{+}-M_{-}}\left\{\Phi^{\pm} e^{\pm i \left(-\omega_{\phi}t+k_{\phi}x\right)} -\Psi^{\pm} e^{\pm i \left(-\omega_{\psi}t+k_{\psi}x\right)}
\right\},
\end{eqnarray}
and
\begin{eqnarray}
    \delta \varphi && = \frac{2M_1M_{+}}{M_{+}-M_{-}}\left\{\Phi^{\pm} e^{\pm i \left(-\omega_{\phi}t+k_{\phi}x\right)} -\Psi^{\pm} e^{\pm i \left(-\omega_{\psi}t+k_{\psi}x\right)}\right\} \nonumber \\
&&-2 M_1 \Phi^{\pm} e^{\pm i \left(-\omega_{\phi}t+k_{\phi}x\right)},
\end{eqnarray}
and then taking their $xx$ and $tt$ derivatives.

\section{Conclusions}\label{sec:concl}

In this work we have analyzed the propagation modes and polarization states of gravitational waves in the GHMPG in its scalar-tensor representation both from the linearization and Newman-Penrose formalisms. Furthermore, we have analyzed the results in light of recent outcomes from a cosmological perspective \cite{Rosa:2024pzo}, in an attempt to qualify the falsifiability of the theory.

In the linearization formalism, we found that the theory exhibits massless tensor modes that uncouple from the other degrees of freedom in the TT-gauge, and two generally massive scalar modes that decouple from each other, but their intertwined relation corresponds to a single degree of freedom, despite being able to excite up to two scalar polarization states, namely the longitudinal and the breathing modes. We found that the masses of the scalar modes depend on the background values of the potential $V$ and its partial derivatives, implying that one can always fine-tune the potential to eliminate the masses of the scalar modes and obtain a speed of propagation equal to the speed of light. 

Moreover, to account for the massive scalar modes, we have resorted to an adaptation of the Eardley-Lee-Lightman formalism which is built from the Newman-Penrose formalism, since the former originally was valid only for nearly null plane gravitational waves. Through this analysis, we have found that, at the linear level, the GHMPG presents the same tensor modes as GR, although the $+$ polarization mode exhibits a sort of squeezing behavior, and that the breathing and longitudinal modes are excited by a single degree of freedom, in agreement with the results from the linearization method. We recall that in modified gravity theories the 00-component of the perturbed metric field needs not to be nonvanishing as it is encompassed in the Bardeen gauge invariant scalars and is associated with the longitudinal polarization mode \cite{Dong:2025ddi}, despite the TT-gauge being applied to the $ij$-components subspace. 

In general, in a cosmological spacetime given by the Friedmann-Robertson-Walker metric, the dispersion relation for the tensor modes can be parameterized by $h_A''+\left(2+\bar{\nu}+\nu_A\right)\mathcal{H}h_A'+\left(1+\bar{\mu}+\mu_A\right)k^2h_A=0$, where the index $A\in \{L,R\}$, standing for the left-hand or right-hand circular polarization modes, respectively, (which are built from the plus and cross tensor modes as $h_A=\frac{1}{\sqrt{2}}\left(h_+\pm i h_{\times}\right)$, the primes denote derivatives relatively to the conformal time, $\mathcal{H}$ is the Hubble parameter in conformal time, and the parameters $\nu_A, ~\mu_A$ label the effects of the parity violation which include the amplitude and velocity birefringences of the GWs, whilst the parameters $\bar{\nu}, ~\bar\mu$ arise from other modifications such as frequency-independent effects that modify the GWs speed and friction, as well as Lorentz-violating terms as frequency-dependent damping and nonlinear dispersion of GWs \cite{ParityViolation1,ParityViolation2}. Moreover, the parameters $\bar{\mu}, ~\mu_A$ are associated with the speed of the gravitational wave, whilst $\bar{\nu}, ~\nu_A$ provides an amplitude modulation to the gravitational waveform. Therefore, we find no amplitude birefringence, as we have not considered a cosmological spacetime on which the GW propagates (we leave this discussion for future work), and concerning possible terms associated with velocity birefringence, which currently are unconstrained by the data \cite{Macarena}, we do not found any as the dispersion relation of the auxiliary field, $\chi_{\mu\nu}$, is the same as the original perturbed metric field, $\delta g_{\mu\nu}$, and is identical to the GR result. Therefore, the origin of the "squeezing" feature that appears in the non-linear regime cannot be identified by the birefringence approach from the linearization method, at least considering a Minkowskian background far from the source. 

To analyze the observational implications of the theory, we have constrained the scalar fields and the scalar potential following recent results obtained through a dynamical system analysis of cosmology and a compatibility with solar system dynamics. Furthermore, we assumed the waves to be approximately planar due to the large distance to their sources. Under these constraints, the scalar fields are exponentially suppressed at late times, consequently causing a suppression of the scalar modes at present times and reducing the differences between the tensor modes to the previously mentioned squeezing behavior in the $+$ polarization. 

These results imply that, for the theory to be simultaneously acceptable from both cosmology and gravitational wave physics perspectives under these assumptions, the squeezing effect is the only possible differentiation with respect to GR. However, because GR is included as a particular limit of the GHMPG, the absence of a detection of the squeezing effect renders the GHMPG unfalsifiable from the point of view of the propagation of gravitational waves. The same conclusion can be traced about the distinction between GR and the metric and Palatini $f(R)$ models, by taking the limits $\varphi=0$ and $\psi=0$, respectively. 

Moreover, we should stress that our analysis finds the highest number of polarization states that can exist in the GHMPG model. In fact, some subclasses of models that obey the observational constraints that we used can have fewer modes. For instance, the $f(R)=R^2$ model has no propagating scalar, vector, or tensor mode whatsoever \cite{Alvarez-Gaume:2015rwa}. Nevertheless, there are more $f(R)$ models which are also degenerate in that sense of absence of degrees of freedom or instability, both at the linearized level on top of a maximally-symmetric background (thus, including Minkowski) \cite{Casado-Turrion:2024esi}. Such systematization is not easily extended to a two-variable function, such as in the GHMPG. Moreover, there are complementary classification systems that resort to NP-quantities, which are the Petrov-Pirani-Penrose \cite{Petrov:2000bs} and the Plebanski classifications \cite{McIntosh:1981}. These classification systems are beyond the scope of the present work.

The GHMPG theory encompasses other gravity models, such as the metric and Palatini $f(R)$ models, as well as non-linear combinations of both. Our results are compatible with the polarization and degrees of freedom found in the literature for those. Moreover, to assess the different polarization states or behaviors, future gravitational wave detectors that may allow for different angles between pairs of interferometers would be required. In fact, current differential-arm detectors, also known as quadrupolar antennas, such as LIGO and Virgo, are less sensitive to scalar modes than to tensor ones (despite this varies with the sky location of the source although not very sharply), and fail to distinguish between the longitudinal and the breathing polarization modes \cite{LIGO:technicalnote}. We also note that in massive scalar-tensor theories, the longitudinal polarization mode is suppressed relative to the breathing one by a factor $(\lambda/\lambda_C)^2$, being $\lambda$ the wavelength of the gravitational wave, and $\lambda_C$ the Compton wavelength of the massive scalar \cite{CMWill}. Nevertheless, attempts to enhance the scalar detection have been proposed \cite{Maggiore:2020}.

\begin{acknowledgments}
We thank Nicola Tamanini, Maxence Corman, Macarena Lagos, David Hilditch, Francisco Lobo, Lara Sousa, and José Fonseca for fruitful discussions. C.G. acknowledges Fundação para a Ciência e Tecnologia (FCT) through Project UID/04650 - Centro de Física das Universidades do Minho e do Porto, and Project 2024.00252.CERN. JLR is supported by the Project PID2022-138607NBI00, funded by MICIU/AEI/10.13039/501100011033 (``ERDF A way of making Europe" and ``PGC Generaci\'on de Conocimiento"). MASP acknowledges support from the FCT through the Fellowship UI/BD/154479/2022, through the Research Grants UIDB/04434/2020 and UIDP/04434/2020, and through the project with reference PTDC/FIS-AST/0054/2021
(“BEYond LAmbda”). MASP also acknowledges support from the MICINU through the project with reference PID2023-149560NB-C21. C.G. and MASP acknowledge the European Commission through COST Action CA21136 – “Addressing observational tensions in cosmology with systematics and fundamental physics (CosmoVerse)”, supported by COST (European Cooperation in Science and Technology).

\end{acknowledgments}


\end{document}